\documentclass[vecphys]{svmult}

\usepackage{makeidx}         % allows index generation
\usepackage{graphicx}        % standard LaTeX graphics tool
\usepackage{multicol}        % used for the two-column index
\usepackage[bottom]{footmisc}% places footnotes at page bottom
\makeindex             % used for the subject index
\begin{document}

\title*{The innermost regions of Active Galactic Nuclei -- from radio to X-rays}
\titlerunning{AGN from radio to X-rays}
\author{Eduardo Ros\inst{1},
Matthias Kadler\inst{2}$^{\rm ,}$\thanks{NASA Postdoctoral Research Associate} \and
Sarah Kaufmann\inst{3} \and
Yuri Y. Kovalev\inst{1,4} \and
Jack Tueller\inst{2} \and
Kimberly A. Weaver\inst{2} 
}
\authorrunning{Ros, Kadler, Kaufmann et al.}
\institute{Max-Planck-Institut f\"ur Radioastronomie, Auf dem H\"ugel 69, D-53121 Bonn, Germany
\texttt{ros@mpifr-bonn.mpg.de}
\and Astrophysics Science Division, Code 662, NASA's Goddard Space Flight Center, Greenbelt Road, Greenbelt, MD 20771, USA
\texttt{mkadler, tueller, kweaver@milkyway.gsfc.nasa.gov}
\and Argelander-Institut f\"ur Astronomie, University of Bonn, Auf dem H\"ugel 71, D-53121 Bonn, Germany
\texttt{kaufmann@astro.uni-bonn.de}
\and Astro Space Center, P. N. Lebedev Physical Institute, ulitsa Profsoyuznaya 84/32, 117997 Moscow, Russia \texttt{ykovalev@mpifr-bonn.mpg.de}
}
%
% Use the package "url.sty" to avoid
% problems with special characters
% used in your e-mail or web address
%
\maketitle
\index{Ros}
\index{Kadler}
% Use the \index{} command to code your author index

\begin{abstract}
Active Galactic Nuclei can be probed by at different regions
of the electromagnetic spectrum: e.g., radio observations 
reveal the nature  of their relativistic jets and their magnetic fields,
and complementarily, X-ray observations give insight into
the changes in the accretion disk flows.
Here we present an overview over the AGN research  
and results from an ongoing multi-band
campaign on the active galaxy NGC\,1052.  Beyond these studies,
we address the latest technical developments and its impact
in the AGN field: the Square Kilometre Array, a new
radio interferometer planned for the next decade, and the
oncoming X-ray and gamma-ray missions.
\end{abstract}

\section{Background}
The standard model for Active Galactic Nuclei (AGN) proposes
that the energy release is produced
by the accretion of matter onto
super-massive black holes (BH)
%, usually at the heart of giant elliptical galaxies 
\cite{fabian99,lobanov06a,lobanov06b}.   
The AGN is powered by the
conversion of gravitational potential energy
into radiation, although the rotational kinetic energy of the BH 
may also serve as an important source of energy 
\cite{koide02,semenov04,komissarov05}. 
A fraction of the matter is ejected via a poloidal magnetic field 
in a jet perpendicular to the accretion disk surrounding the black hole.
%A hot corona can be present in the region next to the base of the jet.

A region of gas with broad emission lines is located close to the
accretion disk.
Narrow-line emitting clouds are present outside the
disk and torus region.  AGN unification models \cite{antonucci93} presume that 
depending on the viewing angle of the torus-disk-jet complex
the observed galaxy appears as a blazar when the jet points
towards the
observer; as an  object like Seyfert 1, Broad-Line-Radio Galaxy or Quasi-Stellar
Object for
intermediate angles; or as a Seyfert 2 or a Narrow-Line-Radio
Galaxy when the jet lies in the plane of the sky.  
For a phenomenological
taxonomy of the AGN zoo, see e.g.\ Table~1.2 in \cite{krolik99}.

The powerful jets observed
commonly in radio-loud AGN \cite{zensus97}
consist of relativistic (shocked) plasma which may extend up to kiloparsec
scales (showing typically extended radio lobes \cite{miley80,bridle84}), much 
larger in size than their host galaxies.  
Jets 
oriented close to the line of sight 
have favourable observing conditions 
due to relativistic
boosting.

Emission from AGN can be observed
throughout the 
electromagnetic 
spectrum: 
The jet synchrotron emission 
can be probed by radio and millimetre
observations. 
The brightest jets emit also in the optical (e.g., \cite{bahcall95}) 
and in X-rays \cite{harris06}.
The thermal emission of the accretion disk and the surrounding torus
are probed both in temperature 
distribution and in morphology by infrared interferometry 
(e.g., \cite{jaffe04}).
The broad and narrow emission line regions are studied by
optical spectroscopy.  
X-ray imaging and spectroscopy probe the corona, the accretion
disk, and the jet radio lobe region at kiloparsec-scales as
well as compact jets in blazars..
Shocked regions in the jet, where the plasma is hotter,
can emit at energies of up to $\ugamma$-rays \cite{mushotzky93}.

Presently, different tools are available for the astronomers to
probe the nature of AGN.  The spectral energy distribution 
%is based on the measurement of flux density through the 
%broad-band electromagnetic
%spectrum.  This broadband spectrum 
can be split into several components
produced  by 
distinct emission mechanisms (synchrotron,
inverse Compton, thermal, etc.) and affected by absorption.
Spectroscopy of the Fe~K$\ualpha$ X-ray emission (the strongest
fluorescent line due to the highest cross section for
the absorption of all iron atoms 
less ionised than Fe$^{+16}$) probes the
relativistic accretion disk.  This line was first detected in
radio-quiet (Seyfert 1 type) galaxies (e.g.,
MCG--6-30-15 \cite{tanaka95}).  The ``louder'' the galaxy
is at radio wavelengths, the weaker the iron line tends to be.
A thermal ``bump" is usually present in the optical and ultraviolet continuum
spectrum.
Spectroscopy of the broad and the narrow line
regions provides information about the pressure of the medium around the jet.
Measurements of the variability of radio sources yield limits on
the size of the emitting regions (from the smallest
timescales of variations).  Radio- and millimetre-wave 
imaging at the highest resolutions
(very-long-baseline interferometry: VLBI) provide resolutions down to
0.1\,milli-arcsec, reaching typically sub-parsec scales.  This shows
the jet structure at the innermost region of the AGN.  At
the highest frequencies, the emission from the jet base (core) is unveiled 
\cite{lee06,lobanov00}.
Measurements of the polarisation
reveal changes in the magnetic fields present at the jet.

%As just mentioned, the access to the core region in AGN is mainly
%based on results from VLBI observations 
%tracking the sub-parsec evolution and the core-jet region.
Single-dish flux-density and spectral monitoring programs probe
absorption effects and the presence of different 
synchrotron-emitting features in the jets.
These are complemented by X-ray monitoring to probe the accretion
disk via spectroscopy and imaging.
These aspects will be expanded in the next sections. First we
give an overview on
the extensive work being performed on AGN at different wavelengths,
then we will describe an ongoing campaign on the
active galaxy NGC\,1052, and finally
we will provide some prospective view of future observations with
the Square Kilometre Array and new X- and $\ugamma$-ray missions.

\section{Observing the multi-waveband sky}

Pioneering work combining VLBI and X-ray observations was 
performed already in the 1970s \cite{marscher79}.  Important landmarks
in this research are, for instance, the combined radio and X-ray observations
on the quasars 3C\,120 \cite{marscher02} ---where the X-ray flux drops
for days to weeks just prior to the ejection of bright features in the jet, 
and PKS\,1510--089 ---where a superluminal ejection in the jet 
occurred immediately after the start of a major 
X-ray and optical outburst in late 2000 \cite{marscher04}. 

AGN surveys are an essential tool for finding and identifying 
appropriate candidates for successful combined X-ray and radio studies.  
Ref.\ \cite{lobanov06a} summarises most of the ongoing surveys
in VLBI, radio monitoring, infra-red, optical, X-ray
and $\ugamma$-ray wavelengths.

VLBI is a well-established technique. Four decades have elapsed 
since the first experiments (e.g., \cite{kellermann68}), and the discovery
of superluminal motions \cite{whitney71,cohen71} took place
thirty years ago.
The %intensive and extensive 
exploration of the 
radio sky on parsec-scales
has been facilitated dramatically
since the construction of the Very Long Baseline Array
(VLBA) in the mid 1990s.  Starting with the regular observations of the VLBA
a program to monitor the jet kinematics of the most prominent 
radio-loud AGN (over 
hundred) in the northern sky was defined and initiated in 1994. 
This was the 2\,cm Survey 
\cite{kellermann98,zensus02,kellermann04,kovalev05}, continued since 2002 
as the MOJAVE program \cite{lister05,homan06} ---the latter including also
linear and circular polarisation monitoring.

X-ray astronomy is also a relatively young science,
experiencing a new revolution with every new generation of X-ray missions.
After 
\textsl{ASCA} (1993--2001, \cite{tanaka94}) 
and \textsl{Beppo}SAX (1996-2003, \cite{boella97}),
the missions
\textsl{Chandra} (since 1999 \cite{weisskopf00}),
\textsl{XMM-Newton} (since 2001 \cite{jansen01}),
and \textsl{Suzaku} (since 2005 \cite{mitsuda06}) constitute 
the state of the art for X-ray 
imaging and spectroscopy at present.  

X-ray emission from the
sources of the 2\,cm~Survey and the MOJAVE samples has been
studied in detail from the available archival data
\cite{kadler05}: \textsl{2cm-X-sample}, 
was established 
by making use of all publicly available archival data from the 
first four missions mentioned above.  
Originally with 50 sources, the sample is being
completed by a \textsl{Swift} program to observe the remaining
83 objects from the MOJAVE sample \cite{kadler07}.
%The statistical studies show that
%radio-loud, core-dominated AGN have a sharp-peaked distribution
%of hard-power-law photon indices with an average value of
%$\Gamma$=1.68. QSOs show the smallest dispersion of $\Gamma$.  The
%photon index of the soft-excess power-law component correlates
%with the apparent speeds at the radio parsec-scale jets.  No
%correlation of $\Gamma_\mathrm{hard}$ with the core-dominance
%or the radio--to--X-ray loudness was found.  From the imaging,
%eight X-ray jets, four distinct X-ray emission knots, 2 hot spots
%and two jet-associated halos are found in 14 of 26 \textsl{Chandra}-observed
%sources.  
%From this multi-wavelength study, %reported in Ref. \cite{kadler05}, 
%some sources have been selected
%for dedicated monitoring combining the VLBI imaging and
%X-ray spectroscopy and
%radio, optical and X-ray flux density monitoring. 
%Some of the sources selected for this multi-wavelength study are
%III\,Zw2, NGC\,1052, 3C\,120, B0458--020, B0716+714, B1038+064, B1222+216, 
%3C\,273, or B1458+718.  Detailed results on these sources will be
%presented elsewhere.  

%The case of the BL\,Lac object B0716+714 is 
%especially interesting: a tentative iron line is detected, being
%visible as excess emission at $\sim$5.8\,keV in March 2002.  If this
%excess is attributed to the Fe~K$\ualpha$, at 6.4\,keV in the rest
%frame, a value for the redshift of this source can be determined
%to $z$=0.10$\pm$0.04, much lower than previously discussed ranges 
%\cite{wagner96}.  Those findings are reported in 
%Ref. \cite{kadler05}.

In the following section we  report on the multi-wavelength
monitoring
campaign on one particular source from the 2cm-X-Sample,
the active galaxy NGC\,1052.

\subsection{NGC\,1052: the key to jet-disk coupling}

The nearby elliptical galaxy NGC\,1052 
can be classified as a radio-loud object
\cite{kadler04}.
It hosts a twin-jet system oriented
close to the plane of the sky (e.g., \cite{vermeulen03b}). 
NGC\,1052 has been classified
as the prototypical low-ionisation nuclear emission region (LINER).  
This source is particularly suited for connecting radio and X-ray observations,
since it shows an edge-on accretion disk, water maser
emission, an obscuring torus (see below), and it hosts 
mildly relativistic
jets that can be probed by VLBI.

%VLBI observations in the 
%mid 1990s show water maser emission along
%the western jet \cite{claussen98}. 
Detailed multi-wavelength
observations at the centimetre range provide
evidence for an obscuring torus covering partially the
western, receding jet \cite{kellermann99,kameno01,kadler04}.  The 
column densities measured in the radio and X-rays have 
comparable values \cite{kadler04b}.
In X-rays the source shows a flat spectrum and a soft excess \cite{weaver99,
guainazzi99,guainazzi00,kadler04b}.  
Sub-parsec imaging of both jets by VLBI from the
2\,cm~Survey/MOJAVE programme \cite{vermeulen03b} revealed 
motions of 0.26\,c.  A new jet feature is ejected every 3--6 months,
correlated with flux density outbursts.

There are indications \cite{kadler05} that the ejection of a new feature
in the jet, estimated to occur in Epoch 2001.0, is associated with the change
of the relativistic line profile from data taken by 
\textsl{Beppo}SAX (\cite{guainazzi00}, reanalysed in \cite{kadler05})
at epoch 2000.03 and by \textsl{XMM-Newton} at epoch 2001.62, where the
line is broadened.  This is the first detection of a highly relativistic
iron line in a radio-loud AGN with a compact radio jet.  The variability
of the iron line and of the fraction of accreted energy that is
channeled into the jet could be related to changes in the
structure of the magnetic field in and above the accretion disk.
%Instabilities in the inner disk could case the accretion flow
%to be irregular \cite{belloni01}.

Given this scenario, we initiated in mid 2005 a multi mission campaign
to track the birth of new VLBI components at the base of the
jet and counter-jet, to compare those with the flux density
monitoring and spectroscopy in radio and X-rays and to establish
cause-effect relationships in a much more confident way than
the accretion-ejection event reported previously \cite{kadler05}.
This campaign includes, in X-rays: a) \textsl{Rossi X-Ray Timing Explorer} 
(\textsl{RXTE}) flux density monitoring at 2-10\,keV: 30 epochs of
2\,ks each, scheduled every three weeks; b) \textsl{Chandra} imaging
and spectroscopy: one deep observation in Sep 2005; c) \textsl{XMM-Newton}
imaging and spectroscopy: one triggered observation in Feb 2006 so far.
The source is also being monitored by the Burst Alert Telescope 
(BAT; see \cite{barthelmy05})
on-board \textsl{Swift} since the beginning of 2005 (see below).
Radio observations include: 
a) $\lambda\lambda$\,13/6/3.6/2.8/2/1.3/0.9\,cm 
dedicated light curves taken by the 100-m radio telescope 
in Effelsberg, with ca 70\,h observations scheduled every three weeks;
b) $\lambda\lambda$\,31/13/7.7/6/3.9/3.6/2.7/2/1.4\,cm light curves
taken by the RATAN-600 and the Univ.\ of Michigan Radio Astron.\ Obs.\
(UMRAO, \cite{aller03}) in the framework of long-term monitoring 
programs; and c) $\lambda\lambda$\,13/7mm Very Long Baseline Array 
imaging, with 18 observing runs of 6\,h each scheduled every six weeks 
(images from the first epochs are presented in \cite{ros06}).

\subparagraph{\textsl{RXTE} Monitoring}
The \textsl{Rossi X-ray Timing Explorer} (\textsl{RXTE}) is 
monitoring NGC\,1052 since mid 2005 with pointings of 2\,ksec 
each every three weeks.  We concentrate on the analysis of 
the data from the PCA detector \cite{jahoda96}. The data were 
reduced with the \textsc{rex} script\footnote{See 
{\tt http://heasarc.gsfc.nasa.gov/docs/xte/recipes/rex.html}} that 
simplifies the reduction of large amount of data, with standard 
criteria for faint sources.  We used the data of PCU\,2 and layer 
1 which provide the best signal-to-noise ratio.  For the spectral 
analysis we used \textsc{xspec} version 11.3. We  restricted the analysis to the
energy range 2--10\,keV and fitted an absorbed power law with 
a fixed value for the Galactic absorption of
$N_\mathrm{H} = 2.95 \cdot 10^{-20} \; \rm{cm}^{-2}$ \cite{kalberla05}.

\begin{figure}
\centering
\includegraphics[width=1.0\textwidth]{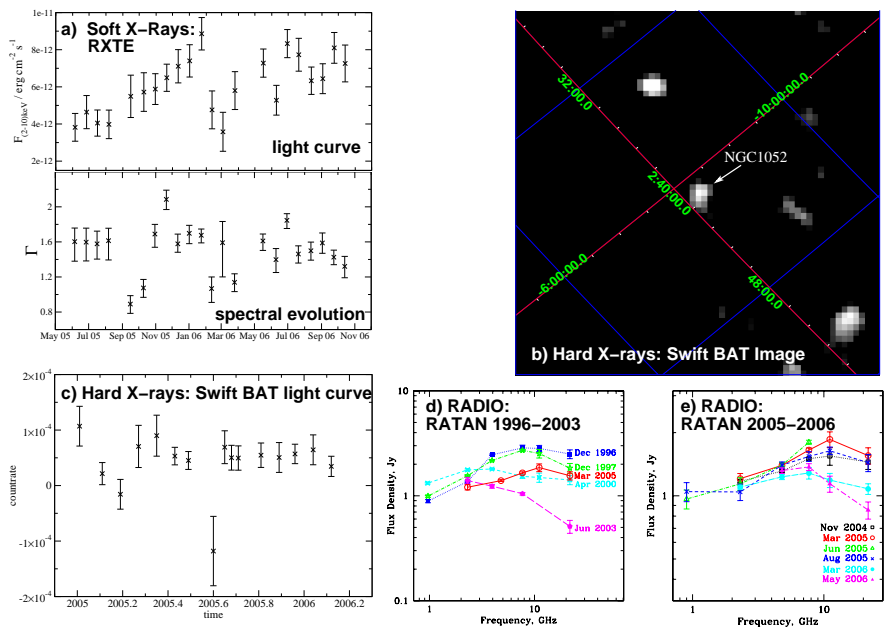}
\caption{
Results from the multi-mission campaign on NGC\,1052.
\textbf{a)} \textsl{RXTE} monitoring results,
showing the 2--10\,keV flux (top) and photon index (bottom) as a function
of time.
\textbf{b)} Hard X-ray image in the 15--150\,keV band. In this $\sim 5^\circ \times 5^\circ$
image, only three sources at $>5 \sigma$ are present; 
\textbf{c)} Hard X-ray light curve from
the first 16 months of BAT observations. 
%Note the different sizes of the error bars, corresponding to different
%exposures, and that negative count rates can statistically occur from the subtraction of the signal from the
%background in weak sources.
\textbf{d)} \& \textbf{e)}
Selected RATAN-600 instantaneous continuum spectra over the last 10
years of observations (d) and during the time of our
multi-frequency campaign (e). The evolution of new radio
flares resulting from parsec-scale jet components is clearly visible
in the overall spectrum changes.
\label{fig:allfigs}
}
\end{figure}

The \textsl{RXTE} light curve (2--10\,keV) and X-ray 
spectral evolution of NGC\,1052 is shown
in Fig.~\ref{fig:allfigs}\,a).
Bearing in mind the complex nature of the X-ray spectrum of NGC\,1052 \cite{kadler05},
the characterisation of the \textsl{RXTE} data from the individual scans with a simple power-law model
requires some extra care in the interpretation. While the flux in the 2--10\,keV band is relatively
insensitive to imperfect spectral modeling, changes of the formal spectral index can be due to changes
of the ratio between various components contributing to the spectrum (e.g., the soft excess, and the
reflection component), as well as due to changes of the primary power-law continuum component or the
absorption. A more careful analysis of the \textsl{RXTE} data with the
detailed spectral composition coming under scrutiny from the deep pointings of {\it XMM-Newton} and
{\it Chandra} is underway.

The \textsl{RXTE} monitoring data show 
a systematic increase of flux 
mid to end 2005 by more than a factor of 2 followed by a
dramatic drop around epoch 2006.1. After that, the source flux rises again over
several months. With the present data, it is difficult to judge whether this
sampled
portion of the light curve corresponds to two outbursts. Alternatively, the
data could be interpreted as showing two long dips similar to 
the ones in 3C\,120
\cite{marscher04} but on longer time scales.
The photon index $\Gamma$ is found to vary most of the time between $1.4$ and $1.8$,
values that are typically seen in Seyfert galaxies. It is interesting to note
that the historically well-known ``unusually flat X-ray spectrum" of NGC\,1052
\cite{weaver99,guainazzi00,kadler04b,kadler04} is found only during two relatively
short time periods in Sep/Oct 2005 and Feb/Mar 2006. These both epochs coincide
with the beginning of a rise in the X-ray flux, indicating that either 
flares occur first at high energies, or vice-versa that the dips last longest
in the soft band.

\subparagraph{\textsl{Swift} BAT Monitoring}
Figure~\ref{fig:allfigs}\,b) shows the hard X-ray 
image (15--150\,keV) of NGC\,1052 from the first 16 months
of \textsl{Swift} BAT monitoring observations \cite{markwardt05}. 
The source is clearly detected with a formal significance of $>$5$\sigma$. 
Unlike the lower-energy bands (radio, soft--medium X-rays),
the BAT light curve (Fig.~\ref{fig:allfigs}\,c) ) shows 
only marginal variability on time scales of months at hard X-rays. Note that
negative count rates can statistically result from the subtraction of two almost equal numbers in the background-dominated limit
and that large error bars indicate periods of
low exposure in the region of NGC\,1052 on the sky. For these two reasons, the negative value at $\sim 2005.6$
should not be over-interpreted. The perhaps more significant feature is the decrease and subsequent increase
in the first three months of 2005. From mid through end 2005, the hard X-ray flux of NGC\,1052 was quasi-constant
within the sensitivity limit of BAT.

The monitoring of NGC\,1052 at hard X-rays will continue throughout the regular all-sky observations of BAT.
This represents a further valuable component in our monitoring campaign of this source. Both long-term trends
and putative higher-amplitude variability on shorter time scales (e.g., due to SSC flares) will be detectable
and can be analysed in view of the variability patterns at lower energies. In particular, it shall be noted that
BAT is sensitive enough and that NGC\,1052 is bright enough in the 15--150\,keV band to detect variability if it
occurs with the same amplitude as in the \textsl{RXTE} band. Thus, even a lack of variability through 2006 in
the BAT light curve would put an important constraint on the nature of the nuclear activity in NGC\,1052 by
attributing most of the variability to a spectral component at soft X-ray energies.

\subparagraph{RATAN-600 Radio Spectra}

We started the long-term monitoring of 1--22~GHz radio spectra of
NGC\,1052 at the 600 meter ring radio telescope RATAN-600 of the Russian
Academy of Sciences in 1996. The observations of continuum spectra are
done almost instantaneously at 1, 2.3, 3.9/4.8, 7.7, 11.1, and 22~GHz
in a transit mode, 2--4 epochs per year. Details of the observations and
data processing can be found in Ref.~\cite{kovalev99}.

Selected RATAN-600 instantaneous continuum spectra over the last 10
years of observations as well as the data accumulated during the time
of our multi-frequency campaign until mid 2006 are presented
in Figure~\ref{fig:allfigs} c) \& d). The flux density at
frequencies over 10\,GHz has dropped by 
a factor of 2 since the
start of the campaign. 
%This factor rises to 5 for variability detected since 1996. 
The observed flaring variability is
most probably due to production and evolution of new parsec-scale features
dominating in the synchrotron spectrum of the compact jet. The flare
spectrum peaks around 10\,GHz in 2005, with SSA radiation below the peak.
The characteristics timescale of the observed long-term radio
variability of NGC~1052 is several months.

\subparagraph{Summary of Some Early Results from the NGC\,1052 Monitoring Campaign}
The monitoring campaign of NGC\,1052 begun in mid 2005 
and
is \textit{en route} to a scrutiny of the jet-disk coupling
in this active galaxy. For one-and-a-half years now, the jet-production 
activity has been monitored at sub-milli-arcsecond angular 
resolution with the VLBA at 22 and 43\,GHz \cite{ros06}. Over this 
period, the source was constantly active in the radio. Its 
radio spectrum has been monitored every three weeks in the cm regime
with the Effelsberg 100-m telescope, and within the long-term 
monitoring programs at the University of Michigan and with 
RATAN-600. We are in the process of 
making a detailed analysis of these data. This will
enable us to compare the jet-production activity to 
accretion-disk probing observations at high energies.

Our preliminary analysis of the first 1.5 years 
of \textsl{RXTE} monitoring data reveals for the first time the
previously missing piece of evidence that the 
2--10\,keV X-ray spectrum of NGC\,1052 is variable on essentially the
same time scales as the radio emission. While a continuation 
of the monitoring
over several variability cycles is important for
quantifying this finding.  The present data support the idea that the 
X-ray and radio components may be directly 
(or indirectly) coupled. In this context, it is important 
to consider our earlier finding that the iron-line 
in NGC\,1052 has varied along with a major jet-ejection event
\cite{kadler05} on exactly these time scales in 
2000/2001. Additional deep X-ray spectroscopic observations
with \textsl{XMM-Newton} and/or \textsl{Suzaku} will likely 
be able to find very different X-ray spectral states in terms
of both the continuum and the iron-line and can be interpreted 
in view of the continuously monitored jet-production activity. 
In particular \textsl{Suzaku}, with its high effective area 
at $\sim 6$\,keV and its broad band pass that covers also the 
hard X-ray regime, will yield important constraints on the iron 
line and will at the same time be able to reconcile the 
\textsl{RXTE} and \textsl{XMM-Newton} results with 
the \textsl{Swift} BAT results.

\section{Instruments for the future}

In this section we discuss the prospectives for AGN research
in the radio and X-rays under the light of the astronomical facilities
planed for this and the next decade.

\subparagraph{Radio: the Square Kilometre Array}

Two new radio facilities will be available in the near future:
operating at very long wavelengths, 
the Low Frequency Array 
in The Netherlands (LOFAR,
\cite{roettgering03}), 
and 
the Atacama Large Millimetre
Array in Chile (ALMA, \cite{brown04}) at the sub-millimetre range.  
In the cm-band, the Square Kilometre Array 
(SKA, \cite{schilizzi05,terzian06,cordes06}),
a new facility reaching 100 times better sensitivity at 
milli-arcsec-resolution is planned for
the next decade.
The scientific
case of the SKA requires a radio telescope with sensitivity
to detect and image atomic hydrogen at the edge of the
universe, which requires a very large collecting area.  The new
instrument should have a fast surveying capability over the
whole sky, which makes a very large angular field of view mandatory.
It should have capability for detailed imaging of the structures
at the sub-arcsecond level, for which a large physical extent
is needed.  Finally, a wide frequency range is needed to
handle the different scientific goals.  The concept which fills
these requirements implies a square kilometre collecting area
in an interferometer array, with a sensitivity two orders of
magnitude larger and a survey speed four orders of magnitude
larger than the Expanded Very Large Array.  The proposed
frequency range should cover from 0.1 to 25\,GHz, with
baselines up to 3000\,km, and a field-of-view of 50 square
degrees at frequencies lower than 1\,GHz.  This project would
cost over one billion euros with a running cost of around
seventy million euros per year.  The so-called Phase 1 of the
project should be ready in 2012, and the complete array
at the end of the 2010s.  The short list of site candidates
favours a location in Southern Africa or Western Australia.
A reference design \cite{skamemo69_06} has been developed recently, including
three system types to be combined in hybrid elements. 
The reference design includes
a sparse aperture array (0.1--0.3\,GHz, a.k.a.\ as
``Era of Recombination" array, similar to LOFAR, providing
wide and multiple independent field-of-views), a
dense aperture array (0.3--1.GHz, a.k.a.\ 
radio ``fish eye" lense with all-sky monitoring capability), 
and a small-dish and ``smart-feed" array  (0.3--25\,GHz,
a.k.a.\ ``radio camera", with $\sim$10-m dishes and 
wide response feeds).  More information on the project
can be found under \verb+http://www.skatelescope.org+.

The SKA will image all radio galaxies in the sky to the
micro-jansky level, especially probing active galaxies in the
radio-quiet regime (e.g., the Seyfert galaxies exhibiting broadened
iron lines in X-Rays).  The project will signify a revolution in
observational cosmology, in the studies of the magnetic universe,
etc., but will also enable a completely new view of the traditional
targets of VLBI research: active galactic nuclei.

\subparagraph{High-energy missions}
Research at high energies will reach 
new frontiers in coming years 
with the missions currently being put into operation or
planned (see \cite{paerels03} for a recent review on X-ray missions):

\textsl{Suzaku} (since 2005) %, \cite{mitsuda06}) 
is specifically designed to study
the Fe~K$\ualpha$ line. Its broad band pass from $0.3$\,keV to $> 100$\,keV
combined with its high effective area at $\sim 6$\,keV and the good
spectral resolution makes it possible to determine the continuum model
and at the same time measure the iron K line. A recent review of
\textsl{Suzaku} observations of iron lines in AGN can be found in
\cite{reeves06}.   

Within its blazar key project, the X-Ray Telescope (XRT) of \textsl{Swift} is currently 
obtaining a large number of X-ray spectra of radio-loud, core-dominated AGN, many of
which have never been observed in the soft X-ray regime since \textsl{ROSAT} in the 
90s and many never above 2\,keV. In particular, \textsl{Swift} is going to complete the
2\,cm-X-Sample of X-ray observed MOJAVE sources \cite{kadler05}, the 133 radio-brightest
compact AGN in the northern sky. \textsl{Swift}
is particularly well suited for such a large pointed-survey program because of its
flexibility. It will greatly enhance our knowledge of radio-loud AGN X-ray spectra by providing
a VLBI-defined statistically complete set of X-ray (and UV) spectra. This will
be particularly valuable to identify more sources similar to NGC\,1052 for combined VLBI and high-energy
studies.

The next planned major step for X-ray observations of AGN will be 
Constellation-X \cite{hornschemeier05}. The current mission design 
of Con-X foresees a sensitivity of $>50$ times better than 
\textsl{XMM-Newton} and \textsl{Suzaku}. The mission will provide the 
highest spectral resolution to date by making use of calorimeter 
detectors.  These will be particularly important for fine-scale studies 
of relativistic broad iron lines in AGN. With Con-X, it will be possible 
to obtain high-quality X-ray spectra of accreting super-massive black-hole 
systems in short snapshot observations.  For sources like NGC\,1052, 
this means that it will be possible to perform high-sensitivity 
accretion-disk monitoring with a minimum of required telescope time. 
At the other end of the electromagnetic spectrum, 
the SKA will for the first time provide the possibility 
to study in detail the radio cores of Seyfert galaxies and other 
radio-quiet AGN. Together, Constellation-X and the
SKA will allow us to perform high-sensitivity combined radio 
and X-ray studies of AGN, both radio-loud and radio-quiet.

In the hard X-ray regime, the all-sky monitoring 
\textsl{Energetic X-ray Imaging Survey Telescope} 
(\textsl{EXIST}, \cite{grindlay03}) would yield a 
sensitivity a factor 50--100 higher than \textsl{Swift}/BAT at the 5--600\,keV.
The science goal of \textsl{EXIST} is the discovery
and study of black holes on all scales from stellar-mass to super-massive
black holes. In the context of hard X-ray blazar studies, \textsl{EXIST} is
expected to boost the number of observationally accessible sources
(e.g., only about 10\% of the MOJAVE blazars are bright enough to be
detected by BAT after 16 months of observations, while
we expect most if not all MOJAVE sources to be easily
detectable by \textsl{EXIST}). For the study of
sources like NGC\,1052 that are bright enough to be studied already
today by the BAT (see above), \textsl{EXIST} will dramatically increase the time
resolution and decrease the minimal detectable variability amplitudes.

%From 2007, a new $\ugamma$-ray facility will be in orbit, the
%\textsl{Gamma-ray Large Area Space Telescope} (\textsl{GLAST}, 
%see \cite{piner05} for a discussion on AGN research with this
%instrument),  
%Different efforts in the radio regime are being triggered now to
%exploit the same targets as this mission: it is planned that new
%$\ugamma$-ray flares trigger VLBI observations, to relate the 
%flaring emission with ejections in the radio jet.  \texttt{MORE DETAILS}

At even higher photon energies, the new $\ugamma$-ray facility \textsl{GLAST}
(Gamma-ray Large Area Space Telescope; \cite{gehrels99}) will be launched
in late 2007. % ([73]). 
It follows in the footsteps of the \textsl{Compton Gamma
Ray Observatory} (\textsl{CGRO}; 1991-1999) whose main instrument EGRET has
discovered that blazars and flat spectrum radio quasars are strong
$\ugamma$-ray emitters \cite{fichtel94}. The third EGRET catalog of
high-energy $\ugamma$-ray sources \cite{hartman99} originally contained
66 high-confidence identifications with blazars. Among the statistically
complete MOJAVE sample of the 133 radio-brightest compact AGN in the
northern sky, 44 have a high-probability EGRET identification according
to revisions of the EGRET-blazar sample 
\cite{mattox97,mattox01,sowards-emmerd03,sowards-emmerd04}).

While the detailed mechanism for the production of gamma-ray emission
has not yet been agreed upon, it is widely accepted that the
$\ugamma$-ray emission from blazars is highly beamed and anisotropic
\cite{dermer94}. It probably takes place close to the central engine
or at the base of the relativistic jet. VLBI observations provide the
best imaging probes close to the central engine, so that one expects to
find differences in the milliarcsecond-scale radio properties
and kinematics between EGRET-detected and not-detected sources. Indeed,
\cite{kellermann04} find that radio jets that are also strong
$\ugamma$-ray sources have faster jets than EGRET undetected sources.
Furthermore, recent work \cite{kovalev05,lister05} indicates that EGRET
sources have more compact radio cores and more highly polarised and more
luminous jet features than non-EGRET sources. In the GLAST era, it will
be possible to investigate these findings in unprecedented detail.

The EGRET counterpart on-board \textsl{GLAST} will be the LAT (Large Area
Telescope) whose capabilities are specifically well-suited for
blazar-variability studies (e.g., \cite{mcenery06}). The highly
superior sensitivity and angular resolution of LAT is expected to result in the
detection of thousands of new $\ugamma$-ray sources, in particular it is
expected that the LAT will be able to monitor the $\ugamma$-ray light
curves of hundreds of bright blazars with high temporal resolution.
Various efforts in the radio regime are being made to exploit the
opportunities offered by this mission: in particular, within the MOJAVE
project, it is planned to trigger VLBI monitoring observations by
\textsl{GLAST}/LAT detected $\ugamma$-ray flares, to relate the flaring activity
with ejections in the radio jet. It will also be possible to use
\textsl{GLAST}/LAT light curves as jet-activity monitors. For sources like
3C\,120 or NGC\,1052, whose mass-accretion can be monitored via X-ray
observations, this will open a new avenue to jet-formation studies.

\begin{small}
\subparagraph*{Acknowledgements} 
%The Very Long Baseline Array is operated by the
%USA National Radio Astronomy Observatory, a facility of the National
%Science Foundation operated under cooperative agreement by Associated
%Universities, Inc.  
%The University of Michigan Radio Astronomy
%Observatory has been supported by the University of Michigan Department
%of Astronomy and the National Science Foundation.
MK was supported by a NASA Postdoctoral Program
Fellowship appointment conducted at the Goddard Space Flight Center.
YYK is a Research Fellow of the Alexander von Humboldt Foundation.
\mbox{RATAN--600} observations were supported partly by the NASA
JURRISS program (W--19611) and the Russian Foundation for Basic
Research (01--02--16812, 02--02--16305, 05--02--17377).
The campaign of observations of NGC\,1052 is being
performed in the framework of
a wide collaboration including 
the 2cm Survey and the MOJAVE teams, and also particularly
E. Ros, E.\ Angelakis, A.\ Kraus, Y.Y.\ Kovalev, A.P.\ Lobanov and J.A.\ Zensus at the MPIfR; 
J. Kerp and S.\ Kaufmann at the AIfA of the University of Bonn; 
M. Kadler, J.\ Tueller and K.\ Weaver at NASA/GSFC; 
A.P.\ Marscher at Boston University; 
and H.D.\ Aller, M.F.\ Aller and J.\ Irwin at the Univ.\ of Michigan.
We thank M.\ Perucho and A.P.\ Lobanov for useful comments to the manuscript.
\end{small}

%%%%%%%%%%%%%%%%%%%%%%%%%%%%%%%%%%%%%%%%%%%%%%%%%%%%%%%%%%%%%%%%%%%%%%  }

%%%%%%%%%%%%%%%%%%%%%%%%%%%%%%%%%%%%%%%%%%%%%%%%%%%%%%%%%%%%%%%%%%%%%%  }

%%%%%%%%%%%%%%%%%%%%%%%%%%%%%%%%%%%%%%%%%%%%%%%%%%%%%%%%%%%%%%%%%%%%%%

\printindex

\begin{thebibliography}{99.}
%
% Use \bibitem to create references.
% Sort the references by alphabetical order
%
% Use the following syntax and markup for your references
%
% Monographs
%\bibitem{monograph} H. Ibach, H. L\"uth: \textit{Solid-State
%Physics}, 2nd edn (Springer, Berlin Heidelberg New York 1996) pp 45--56

% Journal
%\bibitem{journal} S. Preuss, A. Demchuk Jr, M. Stuke et al: Appl. Phys.  A \textbf{61}, 33 (1995)
% Contributed Works
%\bibitem{contribution} D.M. MacKay: Visual stability and voluntary eye
%movements. In: \textit{Handbook of Sensory Physiology}, vol 3, ed by R.
%Jung, D.M. MacKay (Springer, Berlin Heidelberg New York 1973) pp
%307--331

%\bibitem{aharonian04} F. Aharonian, A. Akhperjanian, M. Bellicke et al: A\&A \textbf{421}, 529 (2004)
\bibitem{antonucci93} R. Antonucci: ARA\&A \textbf{31}, 473 (1993)
\bibitem{aller03} M.F. Aller, H.D. Aller, P.A. Hughes: ASP Conf. Ser. \textbf{300}, 159 (2003) %: In \textit{Radio Astronomy at the Fringe}, ASP Conf.\ Ser.\ Vol. 300, ed by J.A. Zensus, M.H. Cohen, E. Ros, (Astron.\ Soc. Pacific, San Francisco, 2003), pp 159--XXX
\bibitem{bahcall95} J.N. Bahcall, S. Kirhakos, D.P. Schneider et al: ApJ \textbf{452}, L91 (1995)
\bibitem{barthelmy05} S.D. Barthelmy, L.M. Barbier, J.R. Cummings et al: Space Science Rev. \textbf{120}, 143 (2005)
%\bibitem{beasley02} A.J. Beasley, D. Gordon, A.B. Peck et al: ApJS \textbf{141}, 13 (2002)
%\bibitem{belloni01} T. Belloni: Ap\&SS Sup. \textbf{276}, 145 (2001)
\bibitem{boella97} G. Boella, R.C. Butler, G.C. Perola et al: A\&AS \textbf{122}, 299 (1997)
%\bibitem{boyle93} B.J. Boyle, R.E. Griffiths, T. Shanks et al: MNRAS \textbf{260}, 49 (1993)
\bibitem{bridle84} A.H. Bridle, R.A. Perley: ARA\&A \textbf{22}, 319 (1984)
%\bibitem{brinkmann97a} W. Brinkmann, W. Yuan, J. Siebert: A\&A \textbf{319}, 413 (1997)
%\bibitem{brinkmann97b} W. Brinkmann, J. Siebert, E.D. Feigelson et al: A\&A \textbf{323}, 739 (1997)
\bibitem{brown04} R.L. Brown, W. Wild, C. Cunningham: Adv. Space Research \textbf{34}, 555 (2004)
%\bibitem{claussen98} M.J. Claussen, P.J. Diamond, J.A. Braatz et al: ApJ \textbf{500}, L129 (1998)
\bibitem{cohen71} M.H. Cohen, W. Cannon, G.H. Purcell et al: ApJ \textbf{170}, 207 (1971)
\bibitem{cordes06} J.M. Cordes: BAAS \textbf{208}, 73.02 (2006) 
\bibitem{dermer94} C.D. Dermer, R. Schlickeiser: ApJS \textbf{90}, 945 (1994)
%\bibitem{donato01} D. Donato, G. Ghisellini, G. Tagliaferri, G. Fossati: A\&A \textbf{375}, 739 (2001)
\bibitem{fabian99} A.C. Fabian: PNAS \textbf{96} 4794 (1999)
%\bibitem{fey96} A.L. Fey, A.G. Clegg, E.B. Fomalont: ApJS \textbf{205}, 299 (1996)
%\bibitem{fey97} A.L. Fey, P. Charlot: ApJS \textbf{111}, 95 (1997)
%\bibitem{fey00} A.L. Fey, P. Charlot: ApJS \textbf{128}, 17 (2000)
%\bibitem{fey04} A.L. Fey, C. Ma, E.F. Arias et al: AJ \textbf{127}, 3587 (2004)
%\bibitem{fey05} A.L. Fey, D.A. Boboltz, P. Charlot et al: ASP Conf. Ser. \textbf{340}, 514 (2005) %. In: \textsl{TITLE HERE}, ASP Conf. Ser. vol. \textbf{340} (ASP, San Francisco, CA, 2005) pp 514-XX
\bibitem{fichtel94} C.E. Fichtel: ApJS \textbf{90}, 917 (1994)
%\bibitem{folkes99} S. Folkes, S. Ronen, I. Price et al: MNRAS \textbf{308}, 459 (1999)
%\bibitem{fomalont00} E.B. Fomalont, S. Frey, Z. Paragi et al: ApJS \textbf{131}, 95 (2000)
%\bibitem{fomalont03} E.B. Fomalont, L. Petrov, D.S. McMillan et al: AJ \textbf{126}, 2562 (2003)
%\bibitem{galbiati05} E. Galbiati, A. Caccianiga, T. Maccacaro et al: A\&A \textbf{430}, 927 (2005)
\bibitem{gehrels99} N. Gehrels, P. Michelson: APh \textbf{11}, 277 (1999)
\bibitem{grindlay03} J.E. Grindlay, W.W. Craig, N.A. Gehrels et al: Proc. SPIE \textbf{4851}, 331 (2003)
\bibitem{guainazzi99} M. Guainazzi, L.A. Antonelli: MNRAS \textbf{304}, L15 (1999)
\bibitem{guainazzi00} M. Guainazzi, T. Oosterbroek, L.A. Antonelli, G. Matt: A\&A \textbf{364}, L80 (2000)
\bibitem{harris06} D.E. Harris, H. Krawczynski: ARA\&A \textbf{44}, 463 (2006)
\bibitem{hartman99} R.C. Hartman, D.L. Bertsch, S.D. Bloom et al: ApJS \textbf{123} 79 (1999)
\bibitem{helmboldt06} J.F. Helmboldt, G.B. Taylor, S. Tremblay et al: AJ, submitted (2006)
%\bibitem{henstock95} D.R. Henstock, I.W.A. Browne, P.N. Wilkinson et al: ApJS \textbf{100}, 1 (1995)
\bibitem{homan06} D.C. Homan, M.L. Lister: AJ \textbf{131}, 1262 (2006)
\bibitem{hornschemeier05} A.E. Hornschemeier, N.E. White, H. Tananbaum: AIP Conf. Proc. \textbf{774}, 383 (2005) %.  In: \textsl{AIP Conf. Proc. 774: X-ray Diagnostics of Astrophysical Plasmas: Theory, Experiment, and Observation}, ed by XXX, (Am. Inst. Phys., Melville, NY, 2005), p 383 
\bibitem{skamemo69_06} International SKA Project Office, Memo 69 \verb+http://www.skatelescope.org/PDF/memos/69_ISPO.pdf+ (2006)
%\bibitem{ishihara04} D. Ishihara, T. Wada, T. Onaka et al: Proc. SPIE, \textbf{5487}, 350 (2004)
\bibitem{jaffe04} W. Jaffe, K. Meisenheimer, H.J.A. R\"ottgering et al: Nature \textbf{429}, 47 (2004)
\bibitem{jahoda96} K. Jahoda, J.H. Swank, A.B. Giles et al: Proc. SPIE \textbf{2808}, 59 (1996)
\bibitem{jansen01} F. Jansen, D. Lumb, B. Altieri et al: A\&A \textbf{365}, L1 (2001)
%\bibitem{jorstad01} S.G. Jorstad, A.P. Marscher, J.R. Wehrle et al: ApJS \textbf{134}, 181 (2001)
\bibitem{kadler04b} M. Kadler, J. Kerp, E. Ros et al: A\&A \textbf{420}, 467 (2004)
\bibitem{kadler04} M. Kadler, E. Ros, A.P. Lobanov et al: A\&A \textbf{426}, 481 (2004)
\bibitem{kadler05} M. Kadler: Compact Radio Cores in AGN: The X-Ray Connection. Dissertation, Rheinische Friedrich-Wilhelms-Universit\"at zu Bonn (2005)
\bibitem{kadler07} M. Kadler et al: in preparation
\bibitem{kalberla05} P.M.V. Kalberla, W.B. Burton, D. Hartmann et al: A\&A \textbf{440}, 775 (2005)
\bibitem{kameno01} S. Kameno, S. Sawada-Satoh, M. Inoue et al: PASJ \textbf{53}, 169 (2001)
\bibitem{kellermann68} K.I. Kellermann, B.G. Clark, C.C. Bare et al: ApJ \textbf{153}, L209 (1968)
\bibitem{kellermann98} K.I. Kellermann, R.C. Vermeulen, J.A. Zensus, M.H. Cohen: AJ \textbf{115}, 1295 (1998)
\bibitem{kellermann99} K.I. Kellermann, R.C. Vermeulen, M.H. Cohen, J.A. Zensus: BAAS \textbf{31}, 856 (1999)
\bibitem{kellermann04} K.I. Kellermann, M.L. Lister, D.C. Homan et al: ApJ \textbf{609}, 539 (2004)
\bibitem{koide02} S. Koide, K. Shibata, T. Kudohy et al: Science \textbf{295}, 1688 (2002)
\bibitem{komissarov05} S.S. Komissarov: MNRAS \textbf{359}, 801 (2005)
\bibitem{kovalev99} Y.Y. Kovalev, N.A. Nizhelsky, Yu.A. Kovalev t al: A\&AS \textbf{139}, 545 (1999)
%\bibitem{kovalev02} Y.Y. Kovalev, Yu.A. Kovalev, N.A. Nizhelsky, A.V. Bogdantsov: PASA \textbf{19}, 83 (2002)
\bibitem{kovalev05} Y.Y. Kovalev, K.I. Kellermann, M.L. Lister et al: AJ \textbf{130}, 2473 (2005)
%\bibitem{kovalev06} Y.Y. Kovalev, L. Petrov, E. Fomalont, D. Gordon: AJ (submitted) \texttt{[arXiv:astro-ph/0607524]} 
%\bibitem{kraus03} A. Kraus, T.P. Krichbaum, R. Wegner et al: A\&A \textbf{401}, 161 (2003)
\bibitem{krolik99} J.H. Krolik: \textit{Active Galactic Nuclei} (Princeton Univ.\ Press, Princeton, NJ 1999), p 20
\bibitem{lee06} S.S. Lee, A.P. Lobanov, T.P. Krichbaum et al. In: \textit{8$^\mathrm{th}$ EVN Symp.}, Proc. of Sci., in press (2006) \texttt{[arXiv:astro-ph/0611308]} 
\bibitem{lister05} M.L. Lister, D.C. Homan: AJ \textbf{130}, 1389 (2005)
\bibitem{lobanov00} A.P. Lobanov, T.P. Krichbaum, D.A. Graham et al: A\&A \textbf{364}, 391 (2000)
\bibitem{lobanov06a} A.P. Lobanov, J.A. Zensus. In: \textsl{Exploring the Cosmic Frontier: Astrophysical Instruments for the 21st Century}, ESO Astrophysical Symposia Series, ed by A.P. Lobanov (ESO, 2005), in press \texttt{[arXiv:astro-ph/0606143]}
\bibitem{lobanov06b} A.P. Lobanov. In: \textit{8$^\mathrm{th}$ EVN Symp.}, Proc. of Sci., in press (2006)
%\bibitem{ma98} C. Ma, E.F. Arias, T.M. Eubanks et al: AJ \textbf{116}, 516 (1998)
\bibitem{markwardt05} C.B. Markwardt, J. Tueller, G.K. Skinner et al: ApJ \textbf{633}, L77  (2005)
\bibitem{marscher79} A.P. Marscher, F.E. Marshall, R.F. Mushotzky et al: ApJ \textbf{233}, 498 (1979)
\bibitem{marscher02} A.P. Marscher, S.G. Jorstad, J.L. G\'omez et al: Nature \textbf{417}, 625 (2002)
\bibitem{marscher04} A.P. Marscher, S.G. Jorstad, M.F. Aller et al: AIP Conf. Ser. \textbf{714}, 167 (2003)%. In: \textit{X-Ray Timing 2003: Rossi and Beyond}, AIP Conf.\ Ser.\ vol 714, ed by P. Kaaret, F.K. Lamb, J.H. Swank (Amer. Inst. Phys., Melville, NY), pp 167--173
\bibitem{marscher05} A.P. Marscher: Mem.S.A.It. \textbf{76}, 13 (2005)
%\bibitem{marshall05} H.L. Marshall, D.A. Schwartz, J.E.J. Lovell et al: ApJS \textbf{156}, 13 (2005)
\bibitem{mattox97} J.R. Mattox, J. Schachter, L. Molnar et al: ApJ \textbf{481}, 95 (1997)
\bibitem{mattox01} J.R. Mattox, R.C. Hartman, O. Reimer: ApJS \textbf{135}, 155 (2001)
%\bibitem{mazin06} D. Mazin for the MAGIC collaboration, ApSS in press \texttt{[arXiv:astro-ph/0609152]}
\bibitem{mcenery06} J. Mcenery: ASP Conf. Ser. \textbf{350}, 229 (2006)
\bibitem{mitsuda06} K. Mitsuda, M. Bautz, H. Inoue et al: PASJ, in press (2006)
\bibitem{mushotzky93} R.F. Mushotzky, C. Done, K.A. Pounds: ARA\&A \textbf{31}, 717 (1993)
%\bibitem{lamer96} G. Lamer, H. Brunner, R. Staubert: A\&A \textbf{311}, 384 (1996)
\bibitem{miley80} G. Miley: ARA\&A \textbf{18}, 165 (1980)
%\bibitem{padovani02} P. Padovani, L. Costamante, G. Ghisellini et al: ApJ \textbf{581} 895 (2002)
\bibitem{paerels03} F.B.S. Paerels, S.M. Kahn: ARA\&A \textbf{41}, 291 (2003)
\bibitem{parmar06} A. Parmar.  In: \textsl{High-Resolution X-Ray Spectroscopy: towards Xeus and Con-X}, ed by G. Branduardy-Raymont \verb+http://www.mssl.ucl.ac.uk/~gbr/workshop2/+
%\bibitem{pearson88} T.J. Pearson, A.C.S. Readhead: ApJ \textbf{328}, 114 (1998)
%\bibitem{piccinotti82} G. Piccinotti, R.F. Mushotzky, E.A. Boldt et al: ApJ \textbf{253}, 485 (1982)
%\bibitem{petrov05} L. Petrov, Y.Y. Kovalev, E. Fomalont, D. Gordon: AJ \textbf{129}, 1163 (2005)
%\bibitem{petrov06} L. Petrov, Y.Y. Kovalev, E. Fomalont, D. Gordon: AJ \textbf{131}, 1872 (2006)
\bibitem{piner05} G. Piner: ASP Conf. Ser. \textbf{340}, 55 (2005) %. In: \textit{Future Directions in High Resolution Astronomy: The 10th Anniversary of the VLBA} ASP Conf Ser vol 340 (Astron. Soc. Pacific, San Francisco, CA, 2005), pp 55-61
%\bibitem{polatidis95} A.G. Polatidis, P.N. Wilkinson, W. Xu et al: ApJS. \textbf{98}, 1 (1995)
\bibitem{reeves06} J.N. Reeves, A.C. Fabian, J. Kataoka et al: Astron. Nachr. \textbf{88}, 789 (2006)
%\bibitem{revnitsev04} M. Revnitsev, S. Sazonov, K. Jahoda, M. Gilfanov: A\&A \textbf{418}, 927 (2004)
\bibitem{roettgering03} H. R\"ottgering: NAR \textbf{47}, 405 (2003)
\bibitem{ros06} E. Ros, M. Kadler.  In: \textsl{Primer Encuentro de la Radioastronom\'{\i}a Espa\~nola}, ed by J.C. Guirado, I. Mart\'{\i}-Vidal, J.M. Marcaide (Servicio de Publicaciones, Universidad de Valencia, Spain 2006) in press \texttt{[arXiv:astro-ph/0608424]}
%\bibitem{sambruna02} R.M. Sambruna, M. Eracleous, R.F. Mushotzky: New Astron. Rev. \textbf{46} 215 (2002)
%\bibitem{sambruna04} R.M. Sambruna, J.K. Gambill, L. Maraschi et al: ApJ \textbf{608}, 698 (2004)
%\bibitem{sazonov06} S. Sazonov, M. Revnivtsev, R. Krivonos et al: A\&A in press \texttt{[arXiv:astro-ph/0608418]}
\bibitem{schilizzi05} R.T. Schilizzi. EAS Pub. Series \textbf{15}, 445 (2005)
\bibitem{semenov04} V. Semenov, S. Dyadechkin, B. Punsly: Science \textbf{354}, 972 (2004)
%\bibitem{soifer87} B.T. Soifer, G. Neugebauer, J.R. Houck: ARA\&A \textbf{25}, 187 (1987)
\bibitem{sowards-emmerd03} D. Sowards-Emmerd, R.W. Romani, P.F. Michelson: ApJ \textbf{590}, 109 (2003)
\bibitem{sowards-emmerd04} D. Sowards-Emmerd, R.W. Romani, P.F. Michelson, J.S. Ulvestad: ApJ \textbf{609}, 564 (2004)
\bibitem{tanaka94} Y. Tanaka, H. Inoue, S.S. Holt: PASJ \textbf{46}, L37 (1994)
\bibitem{tanaka95} Y. Tanaka, A. Fabian, H. Inoue et al: Nature \textbf{375}, 679 (1995)
%\bibitem{taylor94} G.B. Taylor, R.C. Vermeulen, T.J. Pearson et al: ApJS. \textbf{95}, 345 (1994)
%\bibitem{taylor96} G.B. Taylor, R.C. Vermeulen, A.C.S. Readhead et al: ApJS \textbf{107}, 37 (1996)
%\bibitem{taylor05} G.B. Taylor, C.D. Fassnacht, L.O. Sjouwermann et al: ApJS \textbf{159}, 27 (2005)
%\bibitem{teraesranta05} H. Ter\"asranta, S. Wiren, P. Koivisto et al: A\&A \textbf{440}, 409 (2005)
%\bibitem{terashima03} Y. Terashima, N. Iyomoto, L.C. Ho, A.F. Ptak: ApJS \textbf{139}, 1 (2003)
\bibitem{terzian06} Y. Terzian, J. Lazio: Proc. SPIE \textbf{6267} (2006)
%\bibitem{thakkar95} D.D. Thakkar, W. Xu, A.C.S. Readhead et al: ApJS \textbf{98}, 33 (1995)
%\bibitem{valtaoja92} E. Valtaoja, A. Lahteenmaki, H. Terasranta: A\&AS \textbf{95}, 73 (1992)
\bibitem{vermeulen03b} R.C. Vermeulen, E. Ros, K.I. Kellermann et al: A\&A \textbf{401}, 113 (2003)
%\bibitem{wagner96} S.J. Wagner, A. Witzel, J. Heidt et al: AJ \textbf{111}, 2187 (1996)
\bibitem{weaver99} K.A. Weaver, A.S. Wilson, C. Henkel, J.A. Braatz: ApJ \textbf{520} 130 (1999)
\bibitem{weisskopf00} M.C. Weisskopf, H.D. Tananbaum, L.P. Van Speybroeck, S.L. O'Dell: Proc.\ SPIE \textbf{4012}, 2 (2000)
\bibitem{whitney71} A.R. Whitney, I.I. Shapiro, A.E.E. Rogers et al: Science \textbf{173}, 225 (1971)
%\bibitem{wolf03} C. Wolf, L. Wisotzki, A. Borch et al: A\&A \textbf{408}, 499 (2003)
%http://arxiv.org/abs/astro-ph/0610436
%\bibitem{xu95} W. Xu, A.C.S. Readhead, T.J. Pearson et al: ApJS \textbf{99}, 297 (1995)
%\bibitem{york00} D.G. York: AJ \textbf{120}, 1579 (2000)
%\bibitem{yuan98} W. Yuan, W. Brinkmann, J. Siebert, W. Voges: A\&A \textbf{330} 108 (1998)
\bibitem{zensus97} J.A. Zensus: ARA\&A \textbf{35}, 606 (1997)
\bibitem{zensus02} J.A. Zensus, E. Ros, K.I. Kellermann et al: AJ \textbf{124}, 662 (2002)
\bibitem{zensus03} J.A. Zensus, E. Ros, M. Kadler et al: ASP Conf. Ser. \textbf{300}, 27 (2003) %.: The 2\,cm VLBA Survey. In: \textit{Radio Astronomy at the Fringe}, ASP Conf.\ Ser.\ Vol. 300, ed by J.A. Zensus, M.H. Cohen, E. Ros, (Astron.\ Soc. Pacific, San Francisco, 2003), pp 27--34

% Theses

\end{thebibliography}
\end{document}